\documentclass{aa}
\usepackage{graphicx}
\usepackage{color}
\newcommand\sax{{\it BeppoSAX}}
\begin{document}
\title{\sax\ observations of the accretion-powered X-ray pulsar SMC~X-1}
\author{S. Naik$^{1,2}$ \& B. Paul$^2$}
\offprints{S. Naik : sachi@ucc.ie}
\institute{$^1$ : Department of Physics, University College Cork, Cork, Ireland\\
$^2$ : Tata Institute of Fundamental Research, Homi Bhabha Road, Colaba, Mumbai, India}
\authorrunning{Naik \& Paul}
\titlerunning{SMC~X-1 with \sax}
\date{Accepted for Publication in A\&A}

\abstract{
We present here results obtained from three \sax\ observations of 
the accretion-powered X-ray pulsar SMC~X-1 carried out during the declining
phases of its 40--60 days long super-orbital period. Timing analysis of the 
data clearly shows a continuing spin-up of the neutron star. Energy-resolved 
timing analysis shows that the pulse-profile of SMC~X-1 is single peaked at 
energies less than 1.0 keV whereas an additional peak, the amplitude of which 
increases with energy within the MECS range, is present at higher energies. 
Broad-band pulse-phase-averaged spectroscopy of the \sax\ data, which 
is done for the first time since its discovery, shows that the energy 
spectrum in the 0.1--80 keV energy band has three components, a soft 
excess that can be modeled as a thermal black-body, a hard power-law 
component with a high-energy exponential cutoff and a narrow and weak 
iron emission line at 6.4 keV. Pulse-phase resolved spectroscopy indicates 
a pulsating nature of the soft spectral component, as seen in a few other 
binary X-ray pulsars, with a certain phase offset with respect to the 
hard power-law component. Dissimilar shape and phase of the soft and 
hard X-ray pulse profiles suggest a different origin of the soft and 
hard components.
}
\maketitle

\section{Introduction}
The bright, eclipsing, accretion-powered binary X-ray pulsar SMC~X-1 was 
first detected during a rocket flight (Price et al. 1971). The discovery 
of X-ray eclipses with the $Uhuru$ satellite established the binary nature 
of SMC~X-1. The pulsar, with a pulse period of 0.71 s (Lucke et al. 1976), 
is orbiting a B0I super-giant (Sk 160) of mass $\sim$ 19 
M$_\odot$ (Primini et al. 1977) with an orbital period of $\sim$ 3.89 days 
(Schreier et al. 1972). Since its discovery, observations with various X-ray 
observatories clearly show a steady spin-up of the neutron star in the 
binary system. This makes SMC~X-1 an exceptional X-ray pulsar in which no 
spin-down episode has been observed (Wojdowski et al. 1998). An observed 
decay in the orbital period with a time scale of 3 $\times$ 10$^{6}$ yr 
(Levine et al. 1993) is interpreted as due to tidal interaction between 
the neutron star and the binary companion. The later is presumed to be 
in the hydrogen shell burning phase of its evolution. A super-orbital 
period of 40--60 days in SMC~X-1, analogous to the well known X-ray 
pulsars Her~X-1 and LMC~X-4, was suggested by Gruber \& Rothschild (1984) 
and was confirmed by recent observations with the $RXTE$/ASM and $CGRO$/BATSE 
(Clarkson et al. 2003). Varying obscuration by a precessing accretion disk 
provides a good explanation for the long term quasi-periodic intensity 
variations. 

Although the continuum energy spectrum of accreting X-ray pulsars is
described by a power-law component with an exponential cutoff (White 
et al. 1983), there are some binary X-ray pulsars which show the 
presence of a soft excess over the extended hard power-law component. 
The soft component is detectable only in pulsars which 
do not suffer from absorption by material along the line of sight (Paul et 
al. 2002 and references therein). Pulsations in the soft spectral component 
with a certain phase difference with respect to the hard component are also 
seen in a few X-ray pulsars (Her~X-1: Oosterbroek et al. 1997, 2000, Endo 
et al. 2000, SMC~X-1: Paul et al. 2002, LMC~X-4: Naik \& Paul 2004). Apart 
from the hard and soft spectral components, iron emission line features are 
also seen in many of the X-ray pulsars. Iron K shell emission lines in X-ray 
pulsars are believed to be produced by illumination of neutral or partially 
ionized material in accretion disk, stellar wind of the high mass companion, 
material in the form of circumstellar shell, material in the line of sight, 
or in the accretion column. Pulse-phase-averaged and pulse-phase-resolved 
spectroscopy, therefore, provide important information in understanding these
systems in more detail.

The hard X-ray spectrum (20--80 keV energy band) of SMC~X-1, obtained from 
the High Energy X-ray Experiment ($HEXE$) observations, was fitted with a 
thin thermal bremsstrahlung spectrum with a plasma temperature of $\sim$ 
14.5 keV (Kunz et al. 1993). Though a pure power law spectrum was rejected, 
a power law component modified with an exponential cutoff also provided a 
good fit to the $HEXE$ data. The broad-band X-ray spectrum (0.2 -- 37 keV) of 
SMC~X-1 was earlier studied by fitting combined spectra obtained from the 
$ROSAT$ and $Ginga$ observations (Woo et al. 1995). The energy spectrum is 
best fitted with a model consisting of a cutoff power-law type component,
soft excess which is modeled as a single blackbody component, and a broad 
iron emission line. Pulsating hard X-rays and a non-pulsating soft X-rays 
were detected from observations made with $HEAO~1$ A-2 and $Einstein$~SSS 
(Marshall et al. 1983). However, pulse-timing analysis of the $ROSAT$ 
and $ASCA$ observations shows clear pulsations of the soft X-rays with a 
pulse profile different to that of the hard component (Wojdowski et al. 
1998). Pulse-phase-resolved spectroscopy of $ASCA$ data in 0.5 -- 10.0 keV 
energy band shows a pulsating nature of the soft component with some phase 
difference compared to the hard X-rays (Paul et al. 2002) as is seen in 
some other binary X-ray pulsars. The nearly sinusoidal single peaked profile 
of the pulsating soft component contrasts with the double peaked 
profile seen at higher energies. As the $ASCA$ GIS spectrometers are not 
sensitive at energies where the soft excess dominates ($<$ 0.6 keV), it is 
interesting to probe the nature of the soft spectral component over the pulse 
period of the 0.7 s pulsar in SMC~X-1 with the \sax\ LECS.

In this paper, we present the broad band X-ray spectrum of SMC~X-1 over three 
decades in energy. We have carried out detailed timing and spectral analysis 
of three observations of SMC~X-1 with the Low Energy Concentrator Spectrometers 
(LECS), Medium Energy Concentrator Spectrometers (MECS) and the hard X-ray
Phoswich Detection System (PDS) instruments of \sax\ in the energy band of 
0.1--80.0 keV during decaying state of the 40--60 days super-orbital period 
of SMC~X-1. To examine nature of the soft excess, pulse-phase-resolved 
spectral analysis has been carried out for the observation with highest 
X-ray intensity. In the subsequent sections we give details of the 
observations, the results obtained from the timing and spectral analysis, 
followed by a discussion on the results obtained from these three \sax\ 
observations.

\begin{figure}[t]
\vskip 6.2 cm
\includegraphics{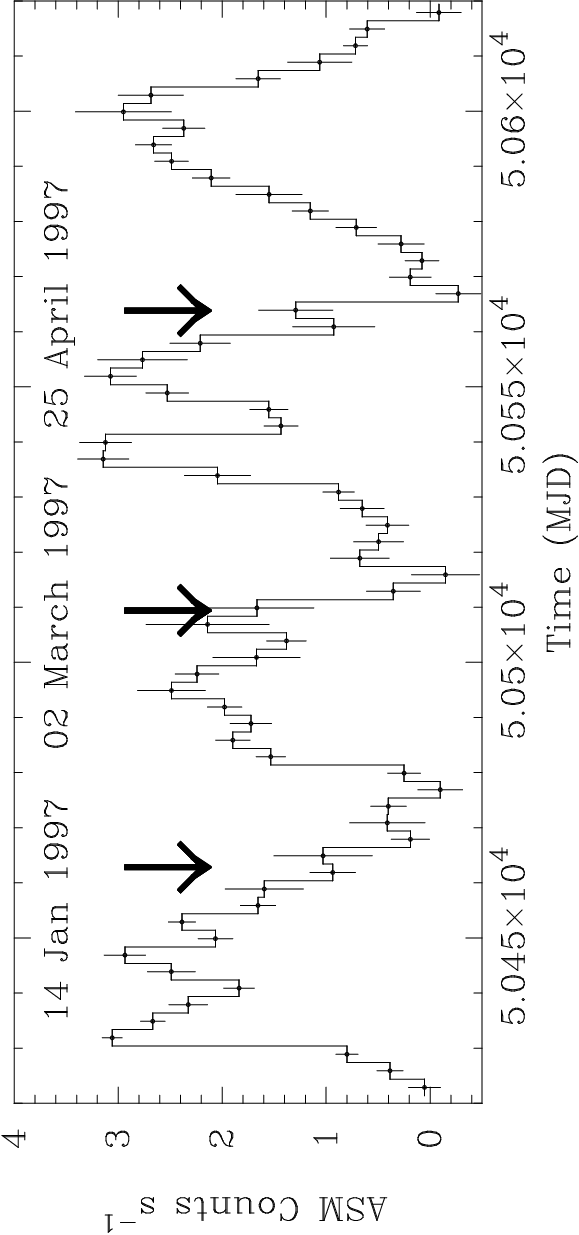}
\caption{The RXTE-ASM light curve of SMC~X-1 from 1996 December 03 (MJD 50420)
to 1997 June 21 (MJD 50620). The arrows mark the dates of the \sax\ 
observations which are used for the analysis.}\label{long}
\end{figure}

\section{Observations}

The long term periodic intensity variation of SMC~X-1, which was
discovered with the instruments on {\it HEAO-I} (Gruber \& Rothschild 1984),
is clearly seen in the {\it RXTE}-ASM light curve. The $RXTE$/ASM light 
curve of the source, from 1996 December to 1997 June (50420 -- 50620 MJD range)
is shown in Figure~\ref{long}. The observations of SMC~X-1 which were made 
with the \sax\ instruments during 1997 January 14 17:26 to 23:55 UT (with
7, 20.5, and 7.5 ks useful exposures for LECS, MECS and PDS respectively), 
1997 March 02 07:14 to 14:30 UT (with 7.5, 21.5 and 7 ks useful exposure) 
and 1997 April 25 16:38 to 23:12 UT (with 7.5, 21.5 and 8.5 ks useful 
exposure) in the high states of 40--60 days quasi-periodic super-orbital 
period, are marked in the figure. All three \sax\ observations were made 
in the orbital phase 0.40 -- 0.54 with the estimated mid-eclipse times 
taken as phase zero. We have used the archival data from these observations 
to study the timing and spectral behavior of the source.

For the present study, we have used data from the LECS, MECS and PDS 
instruments on-board \sax\ satellite. The MECS consists 
of two grazing incidence telescopes with imaging gas scintillation proportional
counters in their focal planes. The LECS uses an identical concentrator system 
as the MECS, but utilizes an ultra-thin entrance window and a driftless 
configuration to extend the low-energy response to 0.1 keV. The PDS detector 
is composed of 4 actively shielded NaI(Tl)/CsI(Na) phoswich scintillators with 
a total geometric area of 795 cm$^2$ and a field of view of 1.3$^o$ (FWHM).
Time resolution of the instruments during these observations was 15.25 $\mu$s 
and energy resolutions of LECS, MECS, and PDS are 25\% at 0.6 keV, 8\% at 6 keV
and $\leq$ 15\% at 60 keV respectively. For a detailed description of the 
\sax\  mission, we refer to Boella et al. (1997) and Frontera et al. (1997).

\begin{figure*}
\vskip 8.5 cm
\begin{center}
\includegraphics{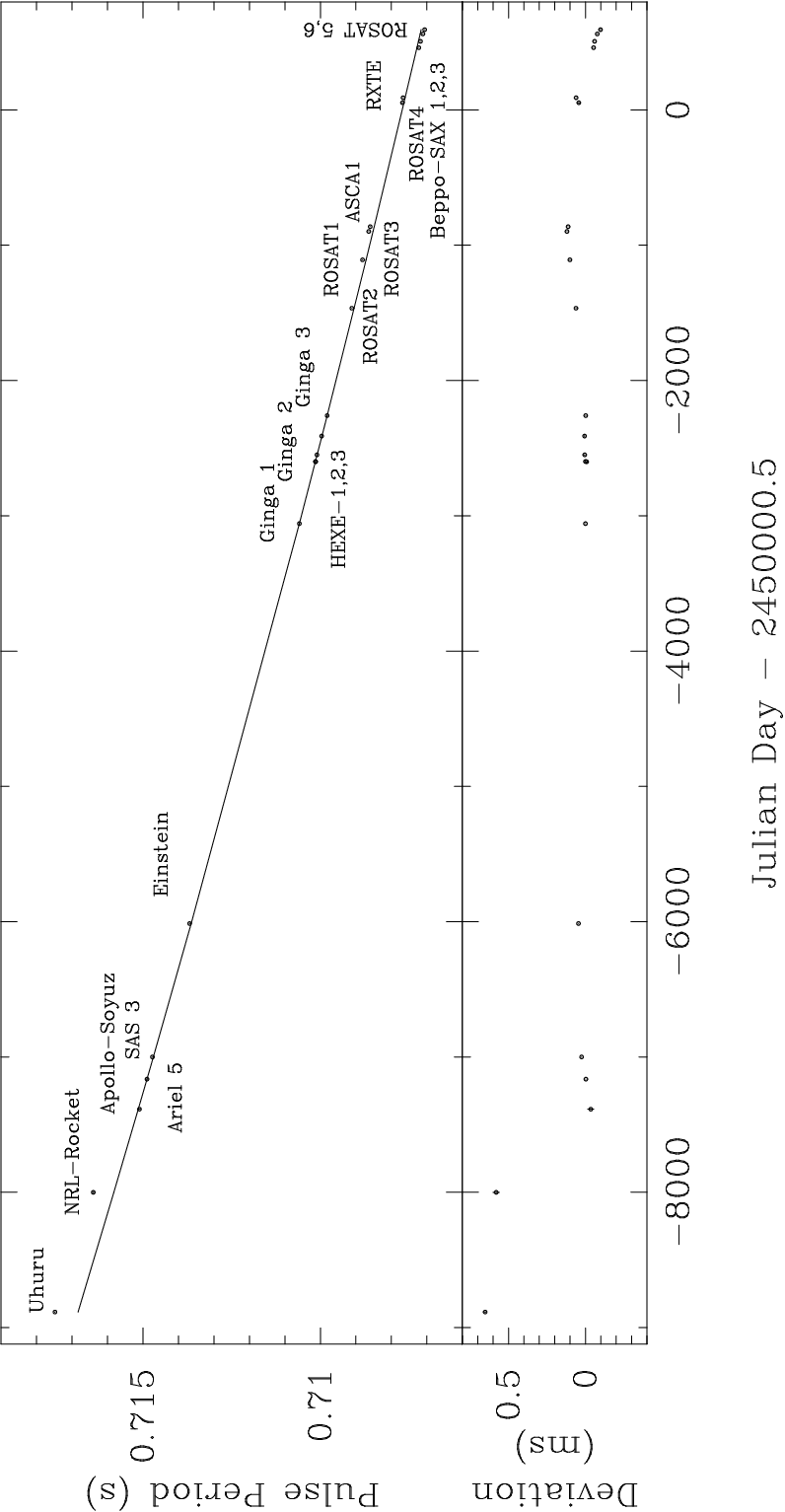}
\caption{Pulse periods of SMC~X-1 obtained from 1997 \sax\ observations along
with that obtained from previous studies are plotted against time together
with a fitted quadratic function (top panel) and the deviations from the fit
(bottom panel).}\label{pph}
\end{center}
\end{figure*}

\section{Timing Analysis}

Data from LECS, MECS, and PDS detectors were used for timing analysis.
The arrival times of the photons were first converted to the same at the 
solar system barycenter. Light curves with time resolution of 7 ms were 
extracted from circular regions of radius $4 \arcmin$ around the source. 
To detect the pulsations during these observations, the light curves were 
first corrected for the arrival time delays due to orbital motion. 
The semi-amplitude of the orbital motion was taken to be 53.4876 ls and 
the mid-eclipse time was derived from the quadratic solution given by 
Wojdowski et a. (1998). After the barycenter correction and correction 
due to the orbital motion, pulse folding and a $\chi^2$ maximization technique 
was applied for pulsation analysis. We have derived the pulse periods of 
SMC~X-1 to be 0.70722816(8) s, 0.70718014(4) s and 0.7071143(7) s on 1997 
January 14, 1997 March 02 and 1997 April 25 respectively. The pulse periods 
obtained from these three \sax\ observations and those obtained from previous 
studies (given in Table~\ref{pph_his}) are plotted against time and shown in 
Figure~\ref{pph} along with a fitted quadratic function of time (upper panel) 
and the deviations from the fit (bottom panel). The coefficients obtained 
from the quadratic fit to the pulse periods are listed in Table~\ref{pph_table}.
It is observed that, since discovery, the neutron star in the SMC~X-1 binary 
system follows a spin-up trend without any spin-down episode.

\begin{figure}
\vskip 12.0 cm
\includegraphics{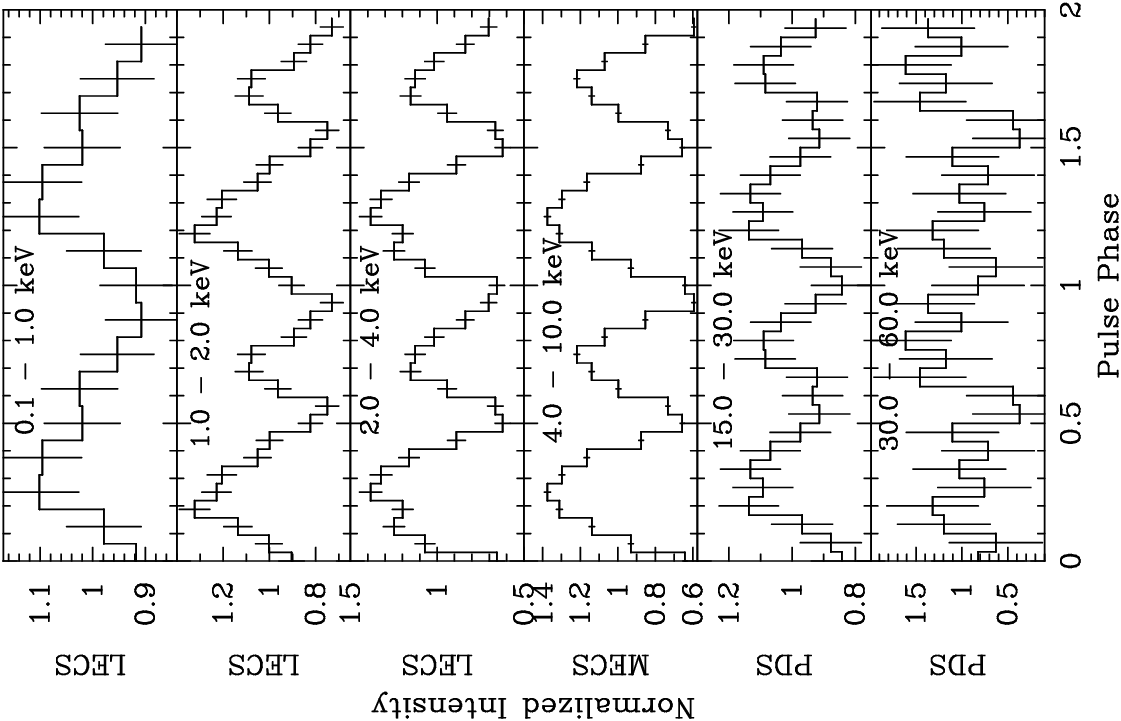}
\caption{The LECS, MECS, and PDS pulse profiles of SMC~X-1 during the high 
state of 40--60 days super-orbital period (1997 March 02 \sax\ observation) are 
shown here for different energy bands with 8 phase bins (top panel) and 16 
phase bins (other panels) per pulse. Two pulses in each panel 
are shown for clarity.}
\label{PP}
\end{figure}

The pulse profiles in different energy bands obtained from the LECS, MECS, and
PDS light curves from the 1997 March 02 observation are shown in 
Figure~\ref{PP}. In 
the low energy band (0.1--1.0 keV of LECS, top panel), the pulse profile is 
nearly sinusoidal and single peaked. However, an additional peak appears in 
the pulse profiles at higher energies as shown in the second, third, fourth, 
and fifth panels of Figure~\ref{PP}. The amplitude of the second peak 
increases with energy and is comparable with that of the main peak as seen 
in 4.0--10.0 keV pulse profile (bottom panel of Figure~\ref{PP}). The light 
curve above 30 keV is mainly background dominated and clear pulsations are
not detected in 30--60 keV energy band.

\begin{table}
\centering
\caption{Summary of pulse period measurements of SMC~X-1 used in present work 
other than that reported in Wojdowski et al. (1998)}
\begin{tabular}{llll}
\hline
\hline
Observatory     &Pulse period (s)  &Epoch (MJD)    &Ref.\\
\hline
\hline
HEXE-1		&0.7101362(15)     &47399.5 &1\\
HEXE-2		&0.7100978(15)	   &47451.5 &1\\
HEXE-3		&0.7099636(15)	   &47591.0 &1\\
\sax-1		&0.7072282(81)	   &50460.9 &2\\
\sax-2          &0.7071801(44)	   &50507.6 &2\\
\sax-3		&0.7071144(71)     &50562.1 &2\\
ROSAT-4		&0.70769(6)	   &50054.0 &3\\
ROSAT-5		&0.707065(10)	   &50593.8 &3\\
ROSAT-6		&0.706707(10)	   &50898.2 &3\\
\hline
\hline
\multicolumn{4}{l}{1 : Kunz et al. (1993), 2 : Present work,}\\
\multicolumn{4}{l}{ 3 : Kahabka \& Li (1999)}\\
\end{tabular}
\label{pph_his}
\end{table}

\begin{table}
\centering
\caption{Coefficient of the quadratic fit to the pulse period data of SMC~X-1}
\begin{tabular}{ll}
\hline
\hline
Parameter$^1$          &Value\\
\hline
\hline
$p_0$		   &0.7076955 s\\
$p_1$              &--9.05 $\times$ 10$^{-7}$ s day$^{-1}$ \\
$p_2$              & 1.38 $\times$ 10$^{-11}$ s day$^{-2}$ \\
$t_0$              &2,450,000.5 JD\\
\hline
\hline
\multicolumn{2}{l}{$P_{pulse} = p_0 + p_1 (t - t_0) + p_2 (t - t_0)^2.$}\\
\end{tabular}
\label{pph_table}
\end{table}

\section{Spectral Analysis}
\subsection{Pulse phase averaged spectroscopy}

For spectral analysis, we have extracted LECS spectra from regions of radius 
$6 \arcmin$ centered on the object (the object was at the center of the 
field of view of the instrument). The combined MECS source counts (MECS~1+2+3) 
were extracted from circular regions with a $4 \arcmin$ radius. The September 
1997 LECS and MECS1 response matrices were used for the spectral fitting. 
Background spectra for both LECS and MECS instruments were extracted from the 
appropriate blank-fields with regions similar to the source extraction regions. 
We rebinned the LECS, MECS and PDS spectra to allow the use of $\chi^2$-statistic.
Events were selected in the energy ranges 0.1--4.0 keV for LECS, 1.65--10.0 keV 
for MECS and 15.0--80.0 keV for PDS where the instrument responses are well 
determined. Combined spectra from the LECS, MECS and PDS detectors, after 
appropriate background subtraction, were fitted simultaneously. All the spectral 
parameters, other than the relative normalization, were tied to be the same for 
all the three detectors and the minimum value of hydrogen column density $N_H$ 
was set at the value of Galactic column density in the source direction.

Broad-band energy spectrum (0.1--80 keV) of SMC~X-1, when simultaneously 
fitted to a single power-law model with line of sight absorption, showed
the presence of an iron emission line at 6.4 keV and significant amount 
of soft excess at low energies ($<$ 1 keV). Paul et al. (2002) found that
the soft excess detected with $ASCA$ spectrometers can be fitted with
different model components such as blackbody emission, bremsstrahlung-type 
thermal component, a soft power-law, or an inversely broken power-law. 
Simultaneous spectral fitting to the LECS, MECS and PDS spectra shows that 
the hard power-law component has an exponential cutoff at $\sim$ 6 keV, as 
was found by Woo et al. (1996), with an e-folding energy of $\sim$ 11 keV. 
With \sax\ combined spectrum, we found that addition of a blackbody emission 
component for the soft excess and a Gaussian function at 6.4 keV for iron 
K$_\alpha$ emission line with the hard power-law continuum model fits the
spectra well. 

The analytical form of the model used for spectral fitting to the 
0.1--80.0 keV band energy spectrum is 
\\
Model--I :\\
\begin{math}
f(E) = e^{-\sigma(E)~N_H} [f_{bb}(E) + f_{pl}(E) f_{cut} + f_{Fe}(E)]
\end{math}
\\
\\
where $E$ is the energy of the incident photon and $\sigma(E)$ is the
photoelectric absorption cross section and \\
\\
\begin{math}
~~~~~~~~f_{bb}(E) = \frac{I_{bb} (E/kT_{bb})^2 (e-1)}{e^{E/kT_{bb}} - 1},~~f_{pl}(E) = I_{pl}~E^{-\Gamma},\\
~~~~~~~~f_{Fe}(E) = \frac{I_{Fe}}{\sqrt{2 \pi \sigma_{Fe}^2}} \mathrm{exp}\left[-\frac{(E - E_{Fe})^2}{2 \sigma_{Fe}^2}\right]\\
~~~~~~~~f_{cut} = \left\{ \begin{array}{ll} 
                    1,                & \mbox{$E < E_c$}\\
                   e^{-(E - E_c) / E_f},  & \mbox{$E \geq E_c$} 
                   \end{array} \right.
\end{math}

\begin{figure}[t]
\vskip 7.3 cm
\includegraphics{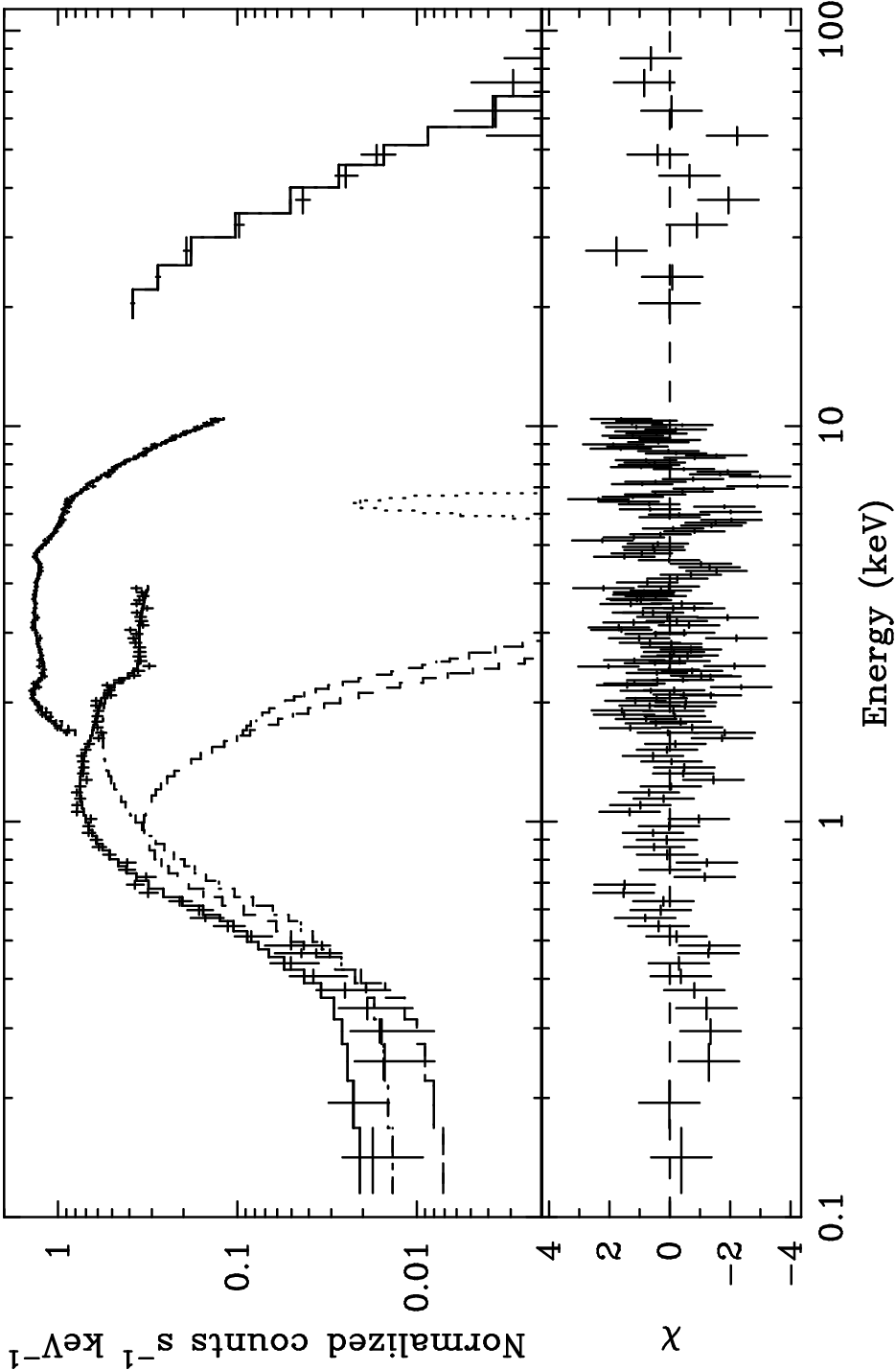}
\caption{Energy spectra of SMC~X-1, during the high intensity state of 40--60
days long super-orbital period, obtained with the LECS, MECS and PDS detectors
of 1997 March 02 observation, along with the best-fit model comprising a soft
blackbody emission, a narrow iron line emission, and a cutoff power law
component. The bottom panel shows the contributions of the residuals to the
$\chi^2$ for each energy bin.}
\label{hss}
\end{figure}

In all three \sax\ observations, the power-law photon index is found to be
$\sim$ 0.8 and hydrogen column density along the line of sight (N$_H$) is in 
the range of 2.1 -- 3.4 $\times$ 10$^{21}$ atoms cm$^{-2}$. The blackbody 
component, of temperature 0.16 -- 0.19 keV, dominates the spectrum at energies 
below 1.0 keV. Iron K$_\alpha$ emission line, centered at 6.4 keV, is found 
to be very weak during all the observations with very low equivalent width. 
Since the iron line center energy is very close to the power-law cut-off 
energy, the presence of the former was separately verified by fixing the 
continuum model based on data outside the 5.5--7.5 keV range. Assuming a 
distance of 65 kpc to SMC~X-1, we have estimated the X-ray luminosity $L_x$ in 
0.1--80 keV energy band to be $5.8 \times 10^{38}$ erg s$^{-1}$, 6.9 $\times 
10^{38}$ erg s$^{-1}$, and 4.6 $\times 10^{38}$ erg s$^{-1}$ for the 
observations on 1997 January 14, March 02, and April 25 respectively. The 
spectral parameters and the reduced $\chi^2$ for all the \sax\ observations 
of SMC~X-1 made in 1997 are given in Table~\ref{spec_par}. From the spectral 
fitting, it can be seen that all the fitted parameters are identical for these 
three \sax\ observations. The LECS, MECS and PDS count rate spectra of 1997 
March 02 \sax\ observation are shown in Figure~\ref{hss} along with 
contributions of individual components in top panel, whereas the residuals 
to the best-fit model are shown in the bottom panel.

\begin{figure}[t]
\vskip 7.3 cm
\includegraphics{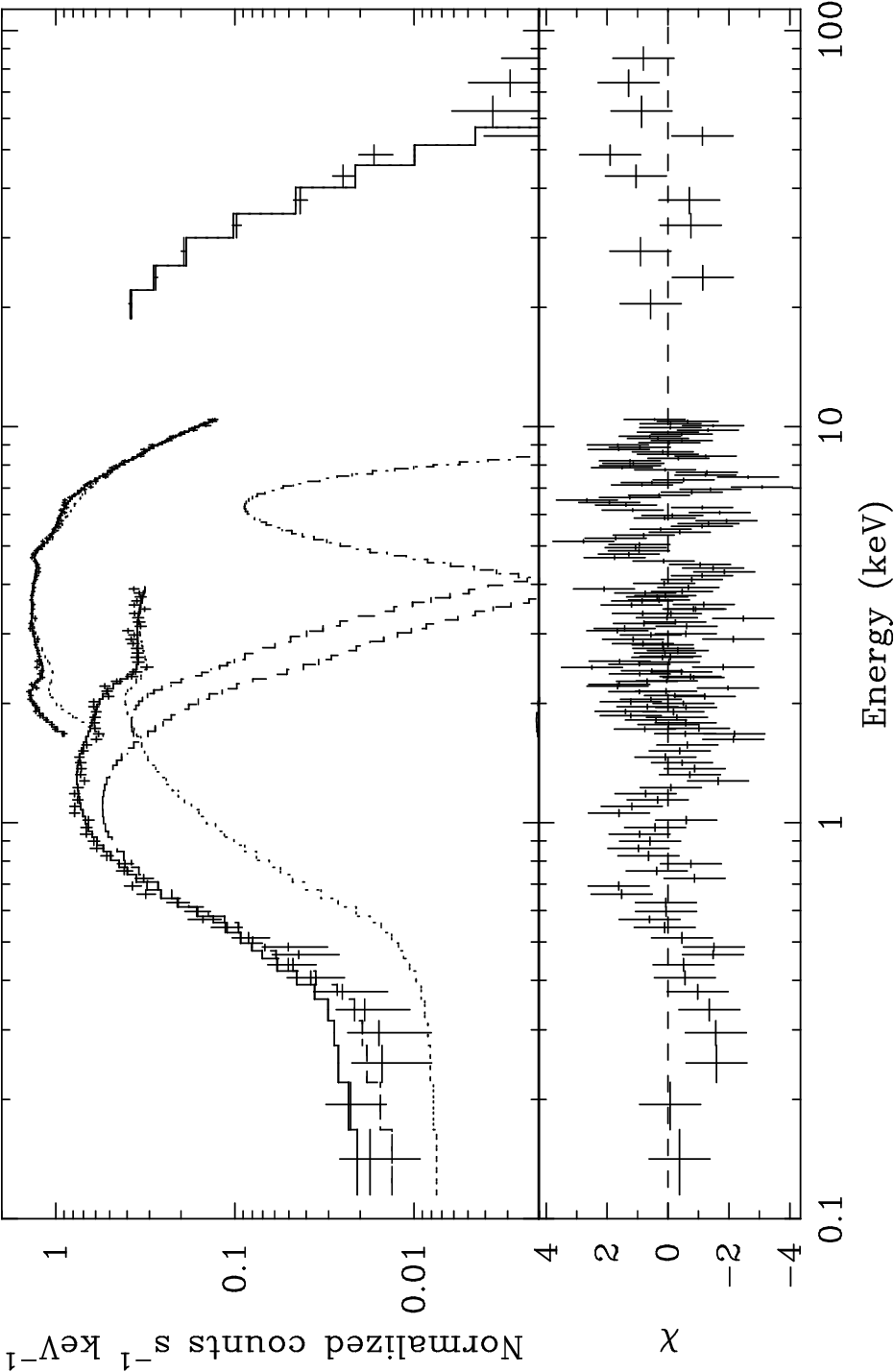}
\caption{Same as Figure~\ref{hss}, but with the Comptonization continuum model.}
\label{ctt}
\end{figure}

We have also tried to fit the LECS, MECS, and PDS spectra with a model 
consisting of a blackbody emission component for the soft excess and the 
Comptonization continuum component (implemented as of ``$comptt$'' in $XSPEC$ 
package) of Titarchuk (1994) as was done for GX~1+4 and RX~J0812.4--3114
by Galloway et al. (2001). The addition of a Gaussian function though improves
the spectral fitting, it gives an unusual emission line, centered around 
5.2--5.7 keV, and a high blackbody temperature of about $\sim$ 2 keV. We have, 
therefore, fixed the iron line energy at 6.4 keV and allowed the width and 
normalization to vary. The spectral parameters of the CompTT model thus 
obtained are also given in Table~\ref{spec_par}. The count rate spectra of 
1997 March 02 observation are shown in the upper panel of Figure~\ref{ctt} 
along with the residuals to the best-fit Comptonization model in the bottom 
panel. It can be seen that the soft spectral component dominates the energy 
spectrum up to $\sim$ 2 keV. Therefore, in this model, we expect that the 
pulse profiles in the energy band of 0.1--2.0 keV should be identical. However, 
the pulse profiles in 0.1--1.0 keV and 1.0--2.0 keV energy bands, as shown 
in Figure~\ref{PP}, are quite different and the reduced $\chi^2$ obtained 
for the Comptonization model is also larger compared to the power-law model 
in all the three observations. We, therefore, conclude that the SMC~X-1 
spectrum is best described as an exponential cutoff power-law with blackbody 
emission (for soft excess) and iron emission line.

\begin{table}
\centering
\caption{Spectral parameters for SMC~X-1 during 1997 \sax\ observations}
\begin{tabular}{llll}
\hline
\hline
Parameter          &14 January        &02 March      &25 April\\
\hline
\hline
\multicolumn{4}{c}{Model -- I (Wabs * (bb + po + Ga) * highecut)}\\
\hline
\hline
N$_H~^1$  	 &3.41$_{-0.46}^{+0.13}$   &2.55$\pm$0.09         &2.11$_{-0.84}^{+1.43}$\\
$\Gamma$         &0.86$\pm$0.02            &0.82$\pm$0.02         &0.77$\pm$0.01\\
$kT$ (keV)       &0.16$\pm$0.01            &0.19$\pm$0.01         &0.19$\pm$0.01\\
$E_{c}$ (keV)    &6.43$\pm$0.11            &6.30$\pm$0.09         &6.41$\pm$0.13  \\
$E_{f}$ (keV)    &11.03$_{-0.15}^{+0.27}$  &10.58$\pm$0.13        &10.14$_{-0.16}^{+0.29}$ \\
$W_0$  (eV)      &25                       &15                    &19\\ 
Total flux$^2$   &4.5$\times$10$^{-10}$    &5.0$\times$10$^{-10}$ &3.1$\times$10$^{-10}$\\
BB flux$^3$	 &1.8$\times$10$^{-11}$    &2.2$\times$10$^{-11}$ &1.1$\times$10$^{-11}$\\
Line flux$^4$    &1.4$\times$10$^{-12}$    &0.9$\times$10$^{-12}$ &0.7$\times$10$^{-12}$\\
Reduced $\chi^2$ &1.2 (176)                &1.2 (168)             &0.98 (167)\\
\hline
\hline
\multicolumn{4}{c}{Model -- II (Wabs * (bb + comptt + Ga))}\\
\hline
\hline
N$_H~^1$	&1.06$^{+0.42}_{-0.31}$  &1.17$^{+0.39}_{-0.40}$  &0.53$^{+0.17}_{-0.11}$\\
$kT$ (keV)	&0.29$^{+0.03}_{-0.03}$	 &0.29$^{+0.03}_{-0.02}$  &0.34$^{+0.03}_{-0.04}$\\
$T_0$ (keV)	&0.81$^{+0.07}_{-0.08}$  &0.95$^{+0.12}_{-0.06}$  &0.94$^{+0.08}_{-0.08}$\\
$kT_e$ (keV)	&5.43$^{+0.21}_{-0.10}$  &5.93$^{+0.19}_{-0.18}$  &5.34$^{+0.16}_{-0.22}$\\
$\tau$		&6.47$^{+0.10}_{-0.23}$  &6.09$^{+0.19}_{-0.34}$  &6.44$^{+0.21}_{-0.22}$\\
Reduced $\chi^2$   &1.3 (169)    &1.3 (169)   &1.1 (169)\\
\hline
\hline
\multicolumn{4}{l}{$\Gamma$ = Power-law photon index, $E_{c}$ = Cutoff energy,}\\
\multicolumn{4}{l}{$E_{f}$ = e-folding energy, $W_0$ = iron equivalent width}\\
\multicolumn{4}{l}{$^1$ : 10$^{21}$ atoms cm$^{-2}$, $^2$ : flux in 0.1--10.0 keV energy }\\
\multicolumn{4}{l}{band in ergs cm$^{-2}$ s$^{-1}$, $^3$ : blackbody flux in }\\
\multicolumn{4}{l}{ergs cm$^{-2}$ s$^{-1}$,  $^4$ : iron line flux in ergs cm$^{-2}$ s$^{-1}$}\\
\end{tabular}
\label{spec_par}
\end{table}

We checked the bremsstrahlung model (Kunz et al. 1993) by fitting the \sax\ 
PDS spectrum of 1997 March observation in 15--80 keV energy band. The plasma 
temperature, obtained from spectral fitting, is found to be 15 keV with a 
reduced $\chi^2$ of 1.7 for 55 degrees of freedom. However, the broad-band 
\sax\ spectrum with LECS, MECS, and PDS data, when fitted simultaneously with
the thermal bremsstrahlung model along with a blackbody emission component for 
soft excess and an iron K$_\alpha$ line, gives a very poor fit with a reduced 
$\chi^2$ of 13.6 for 172 degrees of freedom.

\subsection{Pulse phase resolved spectroscopy}

All the observations of SMC~X-1 with the \sax\ were made near the edge of the 
high state of the 40--60 day super-orbital period (Figure~\ref{long}) and within
0.4--0.54 orbital phase. For pulse phase resolved spectroscopy, we have chosen 
the \sax\ observation made on 1997 March 02 when the source count rate was 
higher compared to other two observations. The data from LECS and MECS 
detectors are used for pulse-phase-resolved spectroscopy as we aimed at 
the study of the nature of the soft spectral component in SMC~X-1. 

\begin{figure}
\vskip 9.5 cm
\includegraphics{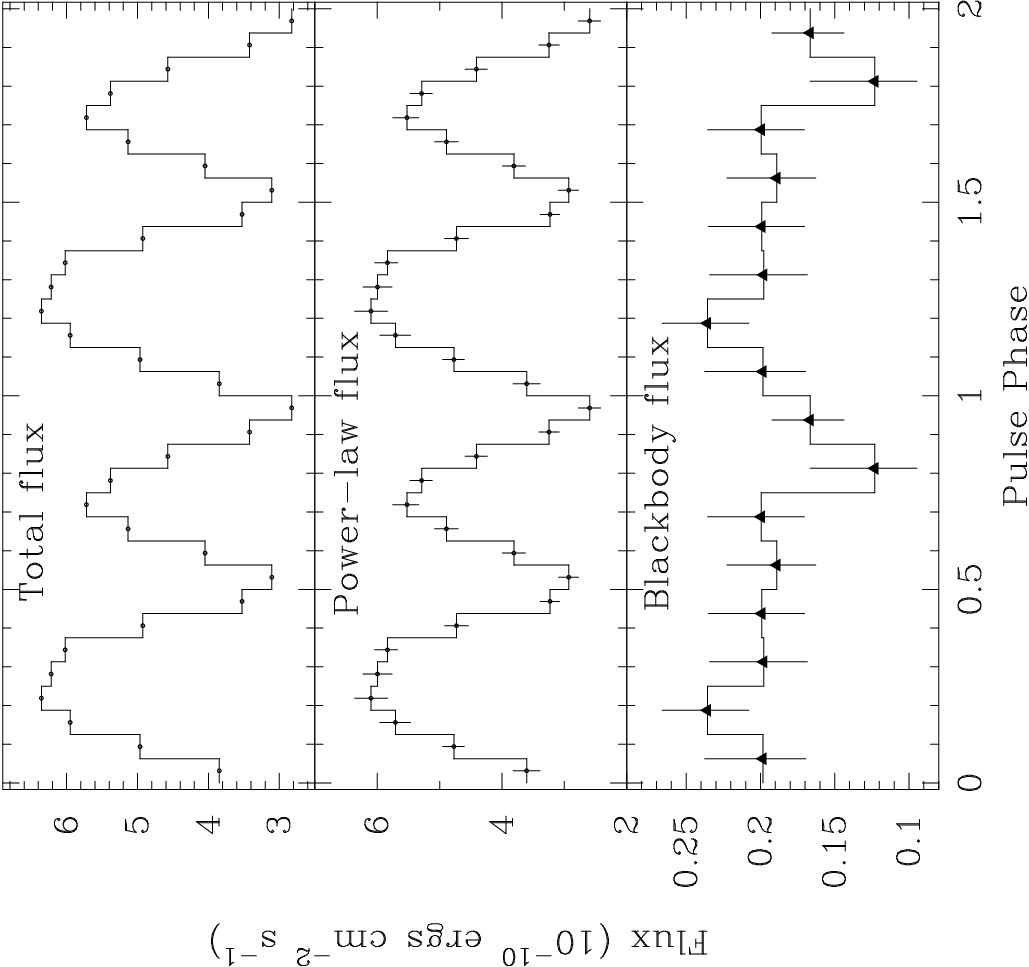}
\caption{Modulation of the total flux, hard power-law flux and flux of the
soft spectral component in 0.1 -- 10.0 keV energy band of SMC~X-1 obtained from
the pulse-phase-resolved spectroscopy of the LECS and MECS spectra obtained
from 1997 March 02 \sax\ observation. The lower panel is rebinned to 8 bins.}
\label{mod}
\end{figure}

Following barycenter and arrival time corrections to the LECS and MECS event
files, spectra were accumulated into 16 pulse phases by applying phase 
filtering in the FTOOLS task XSELECT. As in the case of phase-averaged 
spectroscopy, the background spectra were extracted from source free regions 
in the event files and appropriate response files were used for the spectral 
fitting. The phase-resolved spectra were fitted with a model consisting of 
a high energy cutoff power-law component along with the blackbody emission for
soft excess and iron K$_\alpha$ emission line.  
The iron-line energy, line-width and $N_H$ were fixed to their phase-averaged 
values and all the other spectral parameters were allowed to vary. The 
continuum flux and the fluxes of the soft and hard components in 0.1 -- 10.0 
keV energy range were estimated for all the 16 phase-resolved spectra. The 
modulation in the X-ray flux for the hard and soft spectral components and 
the total flux are shown in Figure~\ref{mod} along with the 1$\sigma$ error 
estimates. Pulse-phase-resolved spectral analysis shows that modulation of 
the the hard power-law flux is very similar to the pulse profile at higher 
energies. The soft spectral component is identical in shape with what was
obtained with the ASCA (Paul et al. 2002). The short duration of the \sax\ 
observation, however, has resulted into relatively large error bars and a 
nonvarying soft excess cannot be ruled out from this data. 

\section{Discussion}

\subsection{Pulse period evolution of SMC~X-1}

Accurate pulse period measurement of a number of X-ray pulsars has been 
achieved over last three decades using various X-ray observatories
(Bildsten et al. 1997). X-ray pulsars which accrete matter from the stellar 
wind of the companion star often show irregular spin rate changes on longer 
time scales, whereas the disk accreting pulsars generally show long-term 
systematic changes in spin period. On the shortest time scales, 
however, the change in spin period appears to be comparable in both the 
groups of X-ray pulsars. 

In the standard accretion-disk model, a pulsar can spin at an equilibrium 
period if the spin-up torque given by the accreting matter is balanced by 
a braking torque due to the interaction of the magnetic field with the 
accretion disk outside the corotation radius (Ghosh and Lamb 1979). For 
a neutron star with given magnetic moment, the equilibrium spin period 
depends on the accretion rate and the pulsar is expected to spin-up or 
down as the accretion rate increases or decreases. The observed correlations 
between the pulse period and the luminosity of X-ray pulsars establish the 
consistency of the model. Using hydromagnetic equations, Ghosh \& Lamb 
(1979) calculated the torque on the neutron star and found that for 
sufficiently high stellar angular velocities or sufficiently low mass 
accretion rates the rotation of the star can be braked while accretion 
continues. 

The observed mean spin-up rate in SMC~X-1 system makes it unique among the 
close binary systems with supergiant companion in which mass accretion takes
place from stellar wind. The period evolution of SMC~X-1 is quite different 
from other persistent HMXB pulsars. $BATSE$ observations (Bildsten et al. 1997)
showed that accreting pulsars with massive companions (eg. Cen~X-3) 
show short term spin-up and spin-down episodes. Though Cen~X-3 
shows 10--100 day intervals of steady spin-up and spin-down trend at a much 
larger rate, it also shows a long term spin-up trend which is the average 
of the frequent transition between spin-up and spin-down episodes (Finger 
et al. 1994). In SMC~X-1, however, the absence of spin-down (torque 
reversal) episodes in more than three decades makes it different from 
other pulsars which show long term spin-up trend. The monotonous decrease 
in the derived pulse period of SMC~X-1 with time suggests that the accretion 
flow has never slowed enough to allow any breaking in the neutron star rotation,
and that SMC~X-1 is far from an equilibrium rotator. It can be noted that 
considering the low metallicity of the SMC/LMC, the wind of the supergiant 
companions alone cannot account for the large persistent X-ray luminosities 
of the pulsars like SMC~X-1 and LMC~X-4. Roche Lobe overflow as a partial 
accretion mechanism is a distinct possibility in these binary systems, which 
is also probable in Cen~X-3.

\subsection{Broad band X-ray spectrum of SMC~X-1}

Since the detection of X-ray emission from SMC in 1971, the accretion powered
high mass X-ray binary pulsar SMC~X-1 has been observed with many different 
X-ray observatories. Though the pulse phase averaged spectral studies of 
SMC~X-1 have been done in different energy bands using X-ray data from various 
instruments such as 20--80 keV from $HEXE$ observations (Kunz et al. 1993), 
0.2--37 keV from $ROSAT$ and $Ginga$ observations (Woo et al. 1995), 0.1--10 
keV from $Chandra$ observation (Vrtilek et al. 2001), 0.5-10 keV from $ASCA$ 
observations (Paul et al. 2002), broad band X-ray spectral study in 0.1--80 keV 
energy range is reported for the first time here. A thermal 
bremsstrahlung model, used to describe the 20--80 keV hard X-ray spectrum 
(Kunz et al. 1993) is ruled out while fitting the source spectrum in 0.1--80
keV energy range. A Comptonization continuum component, used to describe the
spectrum of a few other accretion powered X-ray pulsars, is also found to be
unsuitable for spectral fitting in comparison to the hard power-law continuum 
component. Simultaneous spectral fitting to the broad band X-ray spectrum of 
the source, therefore, shows significant improvement in understanding the 
accretion processes in the binary system. 

Broad-band pulse-phase-averaged spectroscopy of SMC~X-1 shows the presence 
of a weak and narrow iron emission line with very low equivalent width 
($\sim$ 20 eV) and soft excess above the hard power-law continuum component 
as seen in several accreting pulsars. A detailed and systematic analysis of 
X-ray spectra of SMC~X-1 at different phases of its 40--60 days super-orbital 
period would establish the spectral variations of the source over the third 
period. A correlation between the hard X-ray continuum flux and the iron 
emission line flux, a highly variable nature of the otherwise constant  
iron equivalent width during the low intensity states have been found
in other X-ray binary pulsars with super-orbital period (LMC~X-4 and 
Her~X-1, Naik \& Paul 2003). 

\subsection{Nature of the soft excess}

Accreting X-ray binary pulsars which do not suffer from strong absorption
by the material along the line of sight show soft excess over the hard 
power-law component. Her~X-1 (Endo et al. 2000), SMC~X-1 (Paul et al. 
2002), EXO~053109--6609.2 (Haberl et al. 2003, Paul et al. 2004), and 
LMC~X-4 (Naik \& Paul 2004) are the sources in which the difference in 
the pulse profiles at soft and hard X-ray bands along with the presence 
of a soft component over the dominating hard power-law component are already 
reported. Some of the sources also show  pulsations in the soft component. 
The pulsating nature of the soft blackbody component with a certain phase 
difference compared to the hard component and heterogeneous pulse profiles 
at different energy bands suggest different origin of emission of the 
soft and hard components. Endo et al. (2000) discussed about the origin of 
the soft and hard spectral components in Her~X-1 and suggested that the hard 
power-law component originates from the magnetic poles of the neutron star 
in the binary system whereas the origin of the soft blackbody component is 
believed to be the inner edge of the accretion disk. A blackbody or thermal
bremsstrahlung type emission component fits the soft excess of SMC~X-1 and 
LMC~X-4 (Paul et al. 2002). However, from \sax\ observation of LMC~X-4 in
the high state, Naik \& Paul (2004) have established a pulsating nature of the 
soft component which rules out the bremsstrahlung model.

The soft spectral component derived from the \sax\ observations is 
entirely compatible with the results from the ASCA spectra (Paul et al. 
2002). However, a short exposure of only 7.5 ks with the LECS during this 
SAX observation does not allow us to determine accurately the shape and 
phase of the soft component. Therefore, from the present observation we 
cannot rule out a nonvarying soft excess. 

\section*{Acknowledgments}
We thank an anonymous referee for her/his valuable suggestions which improved 
the content of this paper. We thank Dr. P. J. Callanan for a careful reading
of the paper. The \sax\ satellite is a joint Italian and Dutch 
program. We thank the staff members of \sax\ Science Data Center and 
RXTE/ASM group for making the data public. 

{}

\end{document}